\documentclass[a4paper,11pt]{article}

\usepackage{fourier}
\usepackage{color}
 \usepackage{graphicx}
\usepackage{url}
\usepackage[affil-it]{authblk}
\usepackage{amsmath}
\usepackage{wrapfig}

\usepackage[T1]{fontenc}
\usepackage{times}

\usepackage{multirow}
\usepackage{multicol}
\usepackage{subcaption}

\begin{document}

\title{View Sub-sampling and Reconstruction for Efficient Light Field Compression}

\author{Yang Chen}
\author{Martin Alain}
\author{Aljosa Smolic}
\affil{
 V-SENSE project\\
 Graphics Vision and Visualisation group (GV2)\\
 Trinity College Dublin
}

\date{}
\maketitle
\thispagestyle{empty}

\begin{abstract}
Compression is an important task for many practical applications of light fields. Although previous work has proposed numerous methods for efficient light field compression, the effect of view selection on this task is not well exploited. In this work, we study different sub-sampling and reconstruction strategies for light field compression. 
We apply various sub-sampling and corresponding reconstruction strategies before and after light field compression. Then, fully reconstructed light fields are assessed to evaluate the performance of different methods. Our evaluation is performed on both real-world and synthetic datasets, and optimal strategies are devised from our experimental results. We hope this study would be beneficial for future research such as light field streaming, storage, and transmission.  
\end{abstract}

\section{Introduction}
\begin{figure}[t]
  \vspace{-20pt}
  \begin{center}
    \includegraphics[width=0.9\textwidth]{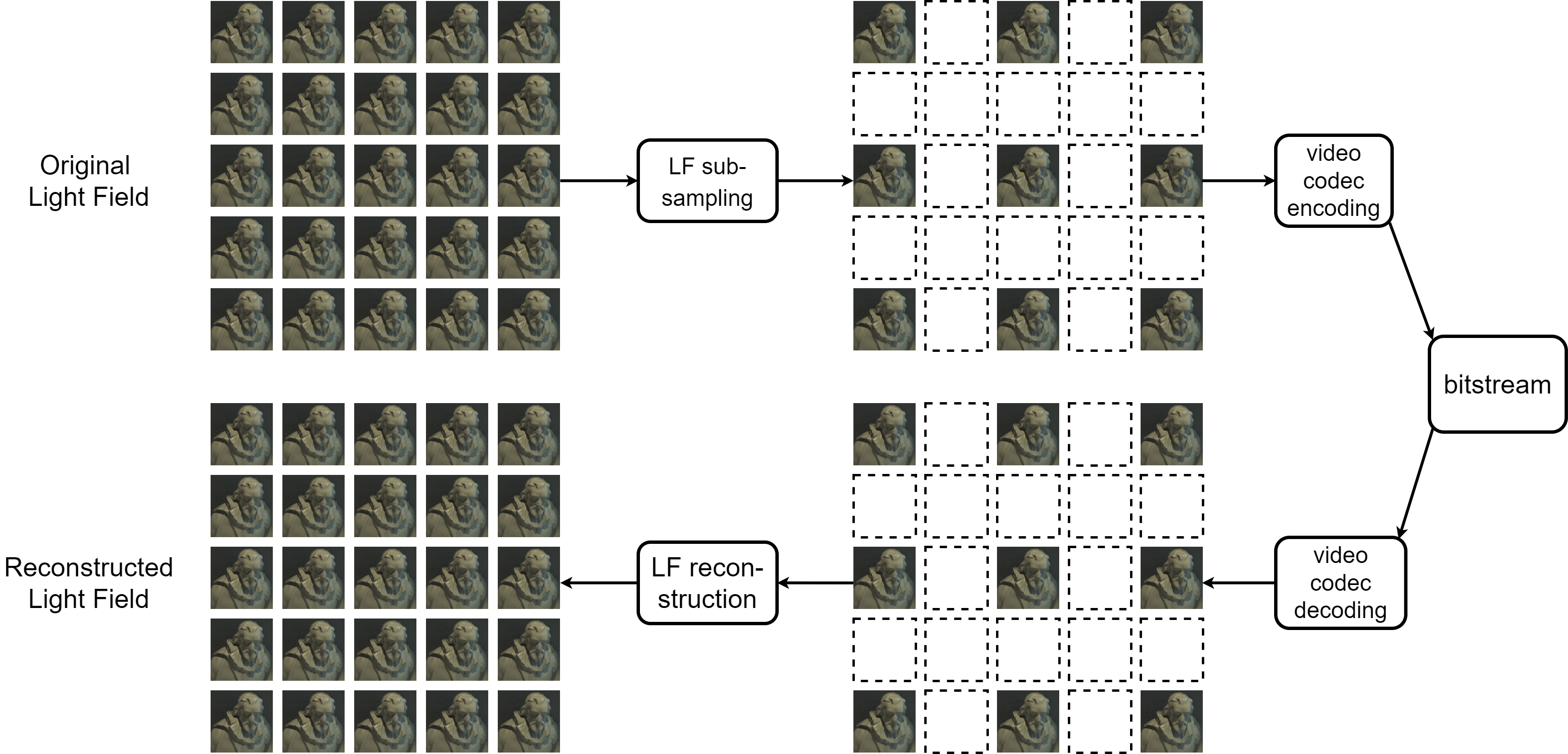}
    \end{center}
  \caption{Light field compression pipeline with sub-sampling and reconstruction strategies.}
  \vspace{-0.4cm}
\end{figure}


A 4D light field is described as a collection of light rays passing through a 3D volume with specific intensity and direction, which can be represented as the interaction between each ray and two parallel planes: image plane and camera plane. The original concept of 4D light fields was firstly introduced by Levoy et. al. in 1996~\cite{Levoy1996}. After years of active research, light fields were applied in various immersive visual applications such as VR, 3DTV and holographic systems. 
The capturing process of the light field usually produces a huge amount of data, which requires plenty of storage space and transmission bandwidth. To reduce these, efficient compression of light field data is crucial for practical applications. One potential way to compress light field data is to sub-sample sub-aperture views for encoding, and to reconstruct the missing views after decoding, which is enabled through advances in light field view synthesis. In this paper, we study different sub-sampling strategies in combination with a recent view synthesis method based on deep learning.

\section{Related Work} 
Light field compression is a crucial research topic, due to the huge amount of data needed for light field imaging. Conti et. al. provided a thorough review about recent light field coding techniques~\cite{conti2020dense}.
One common approach is utilizing light field reconstruction methods to complete view-subsampled light fields, i.e. compressing only a subset of the original views~\cite{chen2017light, zhao2017light, viola2018graph, jiang2017light, jiang2017light2}.
Various methods use a video codec such as HEVC as coding component, while JPEG launched the JPEG Pleno initiative for standardizing compression of plenoptic data including light fields~\cite{ebrahimi2016jpeg}.

Recently, deep learning techniques were introduced to light field compression.
Chen et, al. investigate the impact of subsampling and reconstruction on the light field view synthesis~\cite{chen2020study}.
Chen et. al. proposed a self-supervised learning method to synthesize novel views of light field~\cite{chen2020self}.
Zhao et. al. presented a learning-based method which combines view enhancement and view synthesis to reconstruct complete light fields from decoded views~\cite{zhao2018light}.
Wafa et. al. proposed a deep recursive residual network to synthesize intermediate views after the sparse views are decoded~\cite{wafa2021learning}.
Singh et. al. introduced an end-to-end disparity-aware 3D-CNN for light field compression, which utilizes the disparity information between views and the middle view~\cite{singh2021learning}.
Other works target light field compression using adversarial learning~\cite{jia2018light, bakir2020light, liu2021view}.

All aforementioned approaches perform light field compression with one certain pattern of sub-aperture views, but we argue that the impact of different sub-sampling patterns on compression performance would be worth investigating. Thus, in this paper, we focus on evaluating view selection strategies for light field compression.


\begin{figure}[t]
    \begin{minipage}{.3\linewidth}
        \centering
        \subcaption{\small{row\_2x}}
        \includegraphics[width=0.9\linewidth, scale=.15]{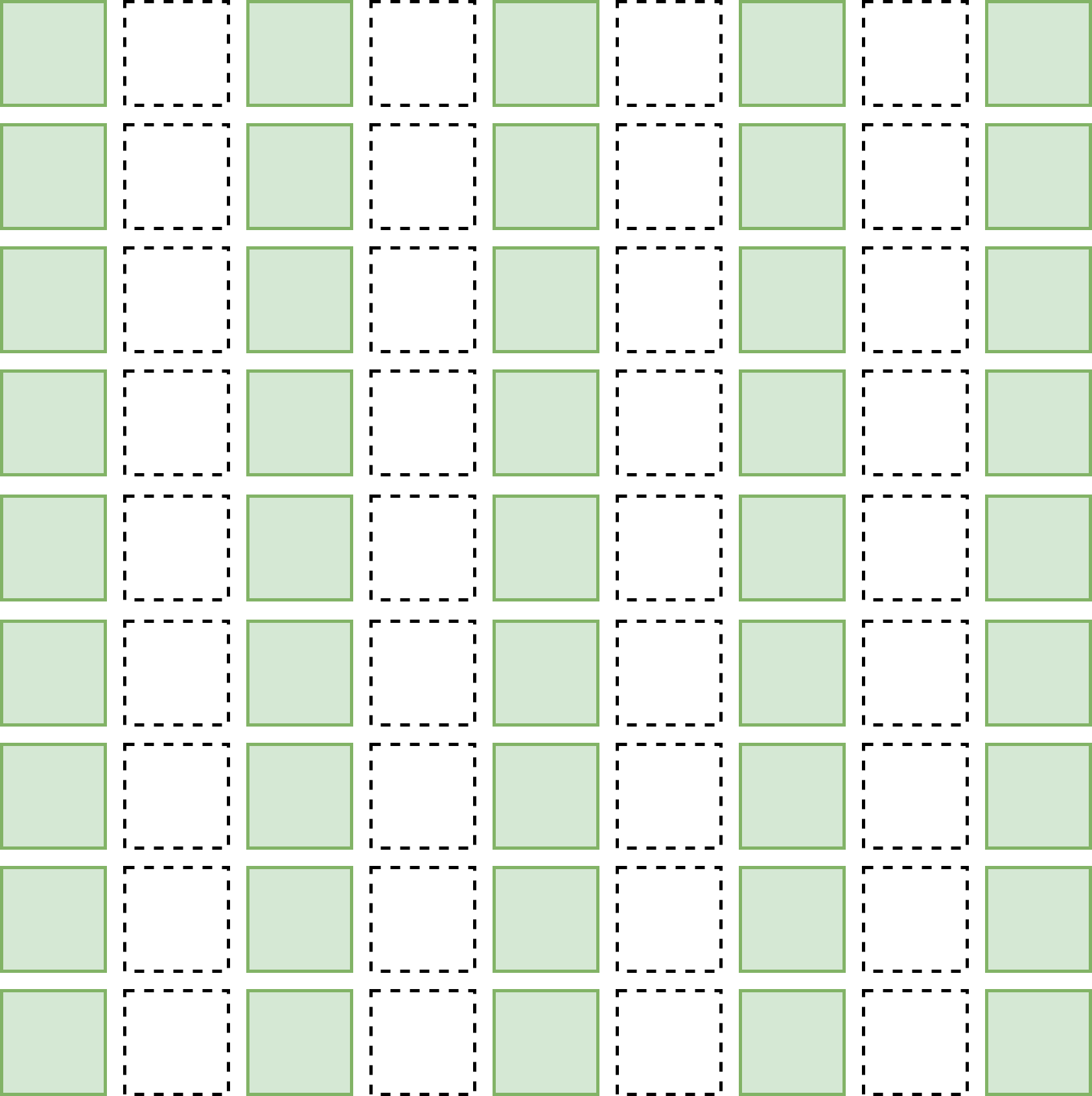}
    \end{minipage}
    \hspace{0.15cm}
    \begin{minipage}{.3\linewidth}
        \centering
        \subcaption{\small{col\_2x}}
        \includegraphics[width=0.9\linewidth, scale=.15]{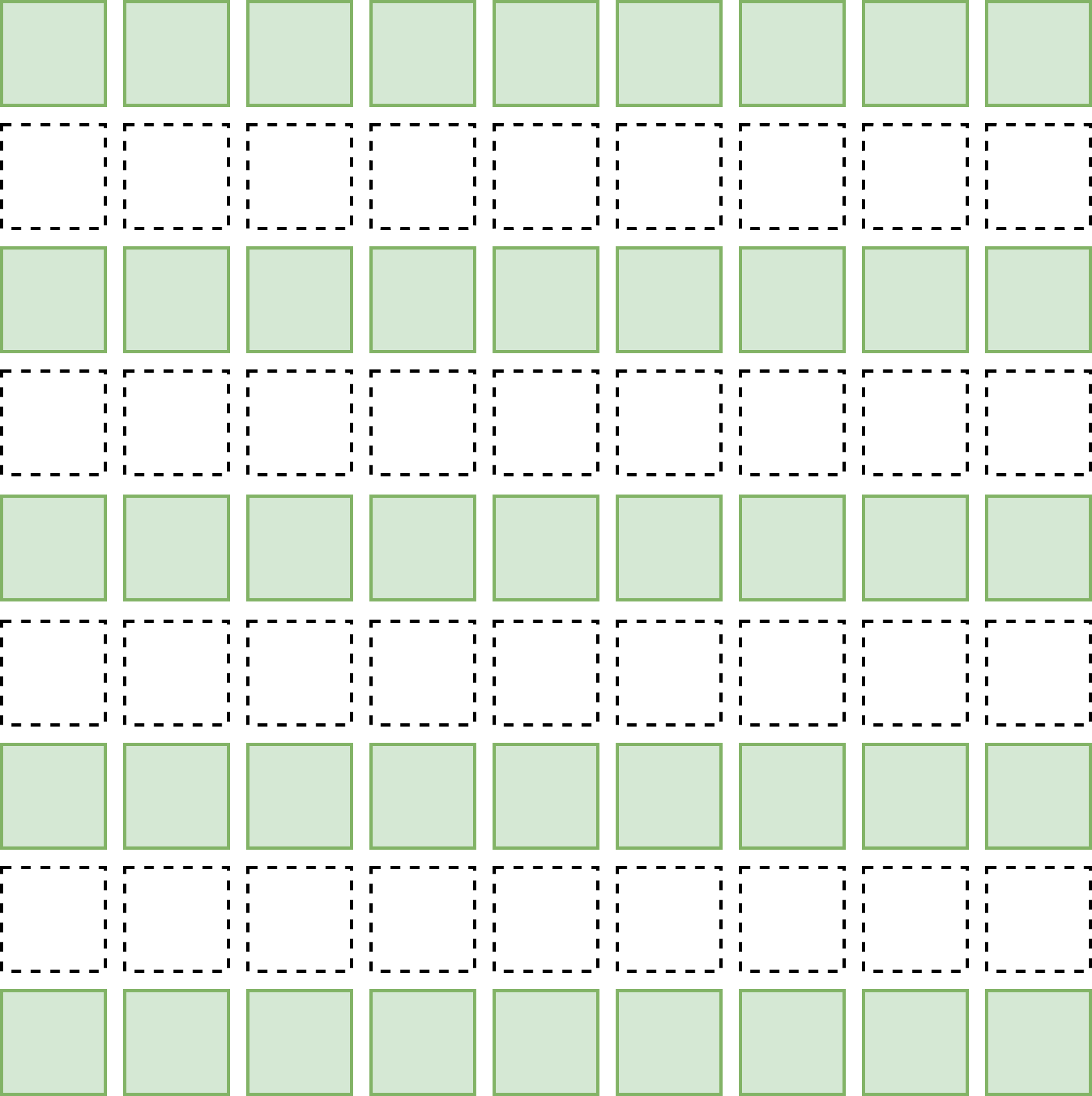}
    \end{minipage}
    \hspace{0.15cm}
    \begin{minipage}{.3\linewidth}
        \centering
        \subcaption{\small{corners\_2x}}
        \includegraphics[width=0.9\linewidth, scale=.15]{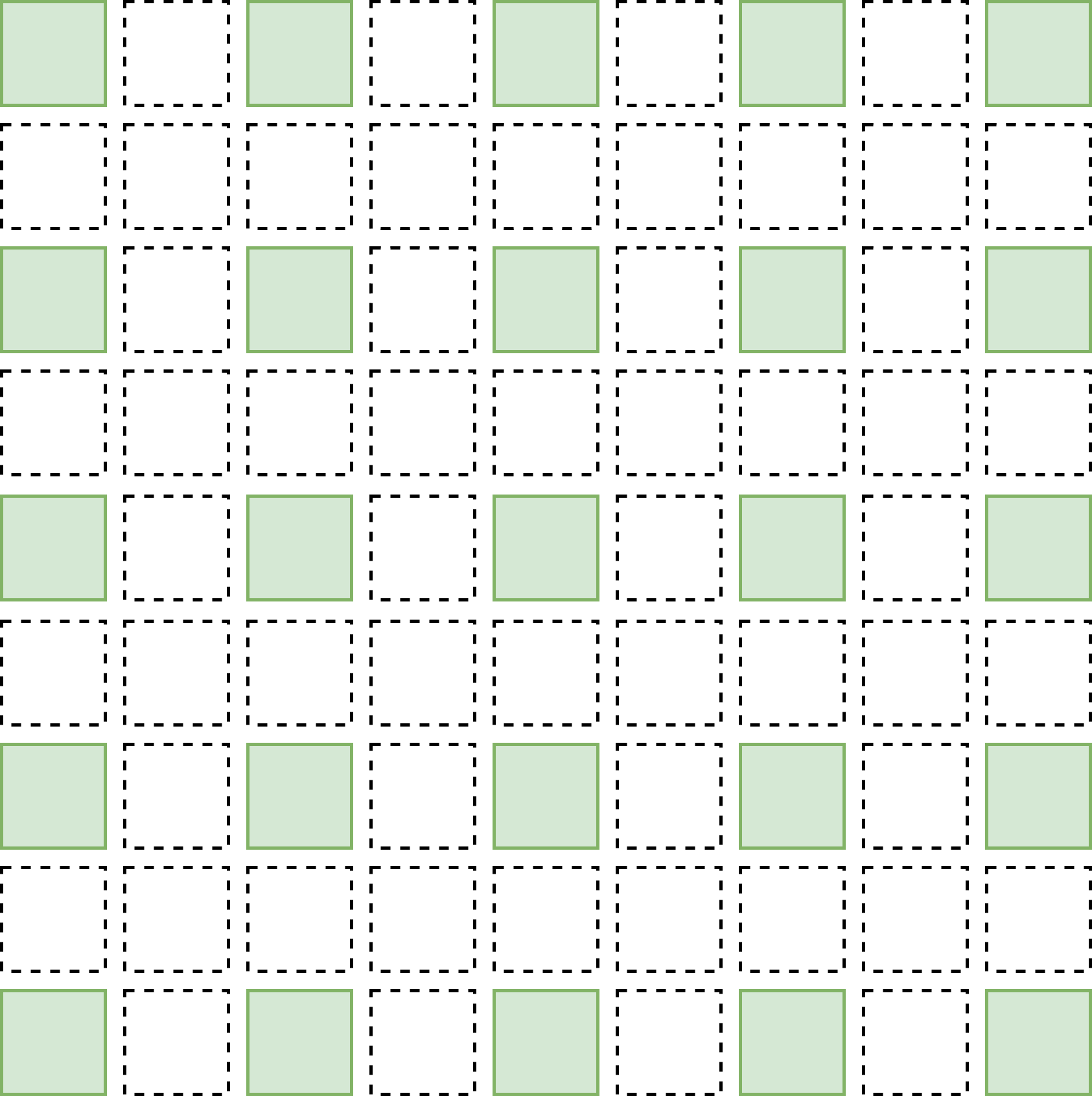}
    \end{minipage}\par\medskip
    \begin{minipage}{.3\linewidth}
        \centering
        \includegraphics[width=0.9\linewidth, scale=.15]{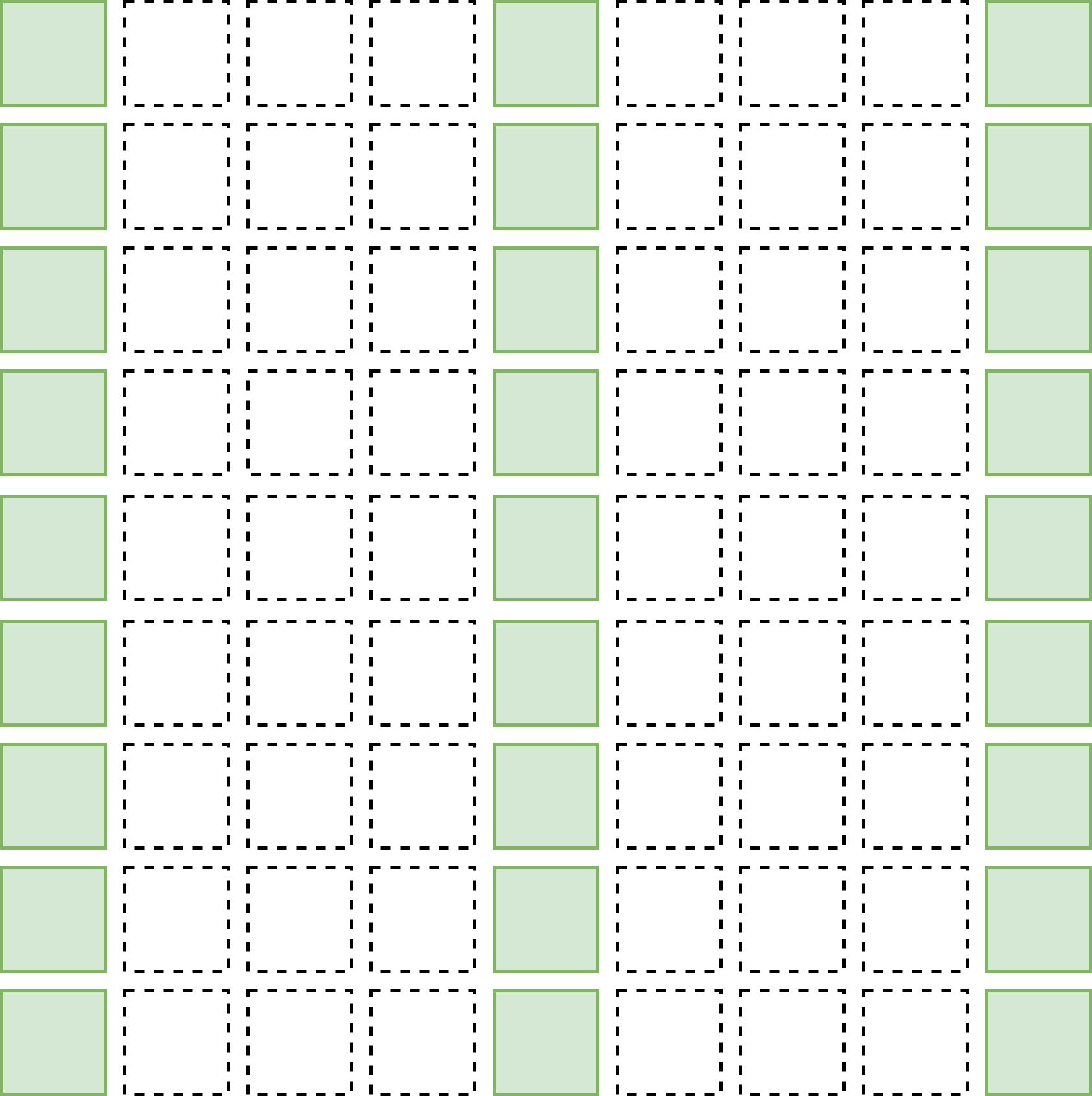}
        \subcaption{\small{row\_4x}}
    \end{minipage}
    \hspace{0.15cm}
    \begin{minipage}{.3\linewidth}
        \centering
        \includegraphics[width=0.9\linewidth, scale=.15]{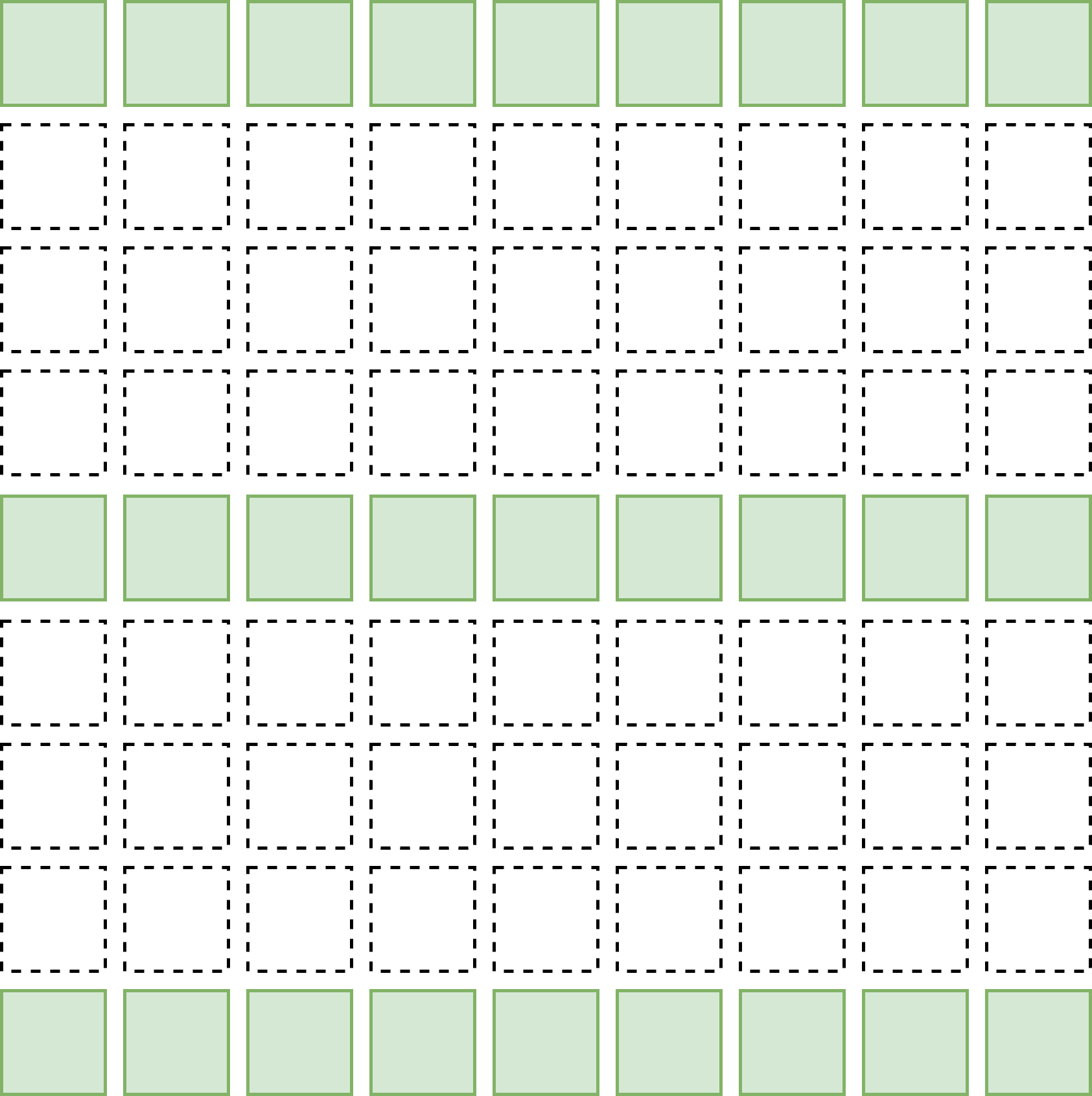}
        \subcaption{\small{col\_4x}}
    \end{minipage}
    \hspace{0.15cm}
    \begin{minipage}{.3\linewidth}
        \centering
        \includegraphics[width=0.9\linewidth, scale=.15]{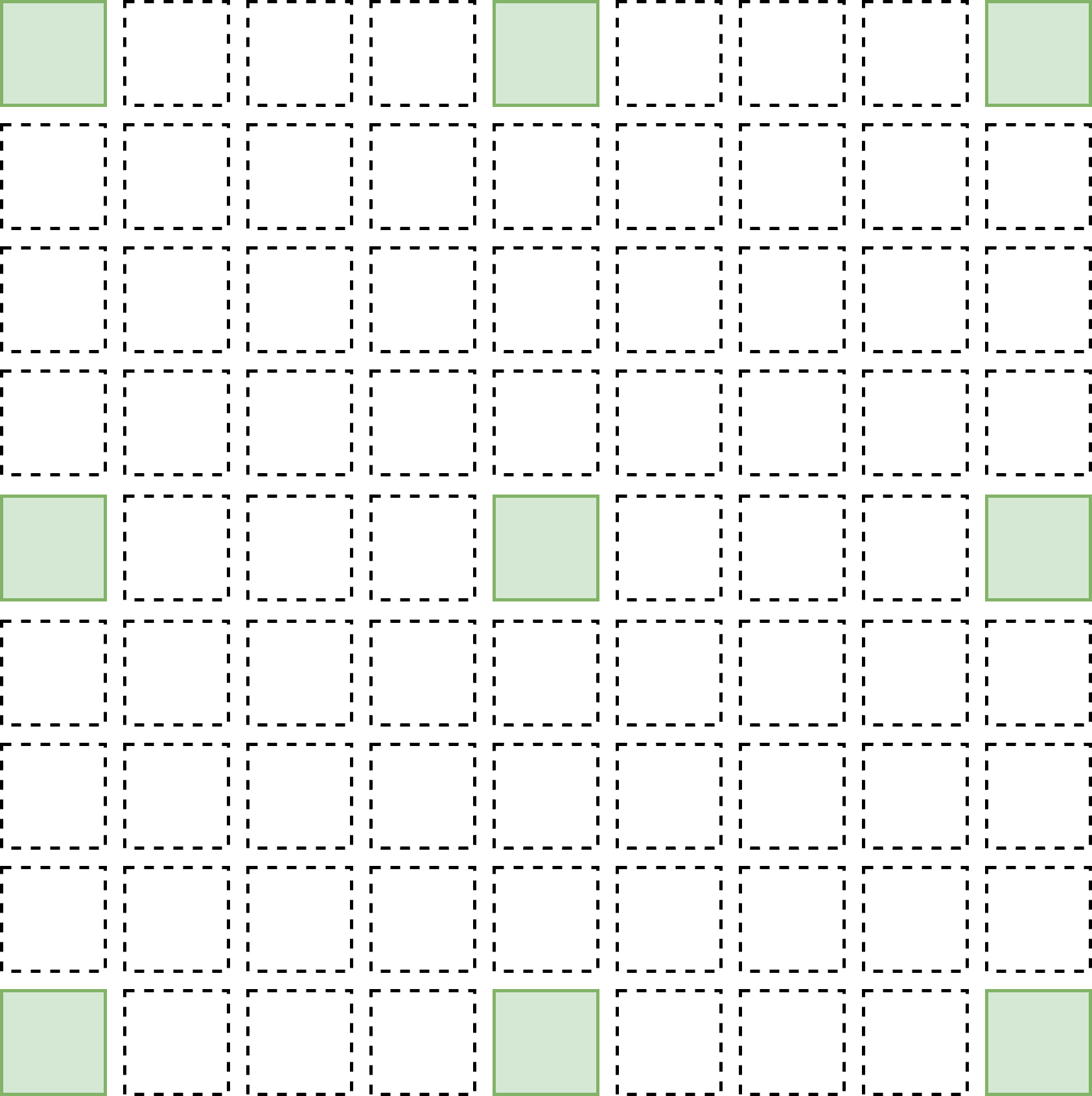}
        \subcaption{corners\_4x}
        \label{subfig:corners_4x}
    \end{minipage}
    \caption{Six strategies to sub-sample views from a $9\times9$ light field. Sub-sampled views are shown in green. The columns show three different type of strategies (row, column and corners). Top (2x) and bottom (4x) rows illustrate different sampling densities.}
	\label{fig:subsampling}
	\vspace{-0.4cm}
\end{figure}

\begin{figure}[!t]
\centering
\includegraphics[width=0.8\linewidth]{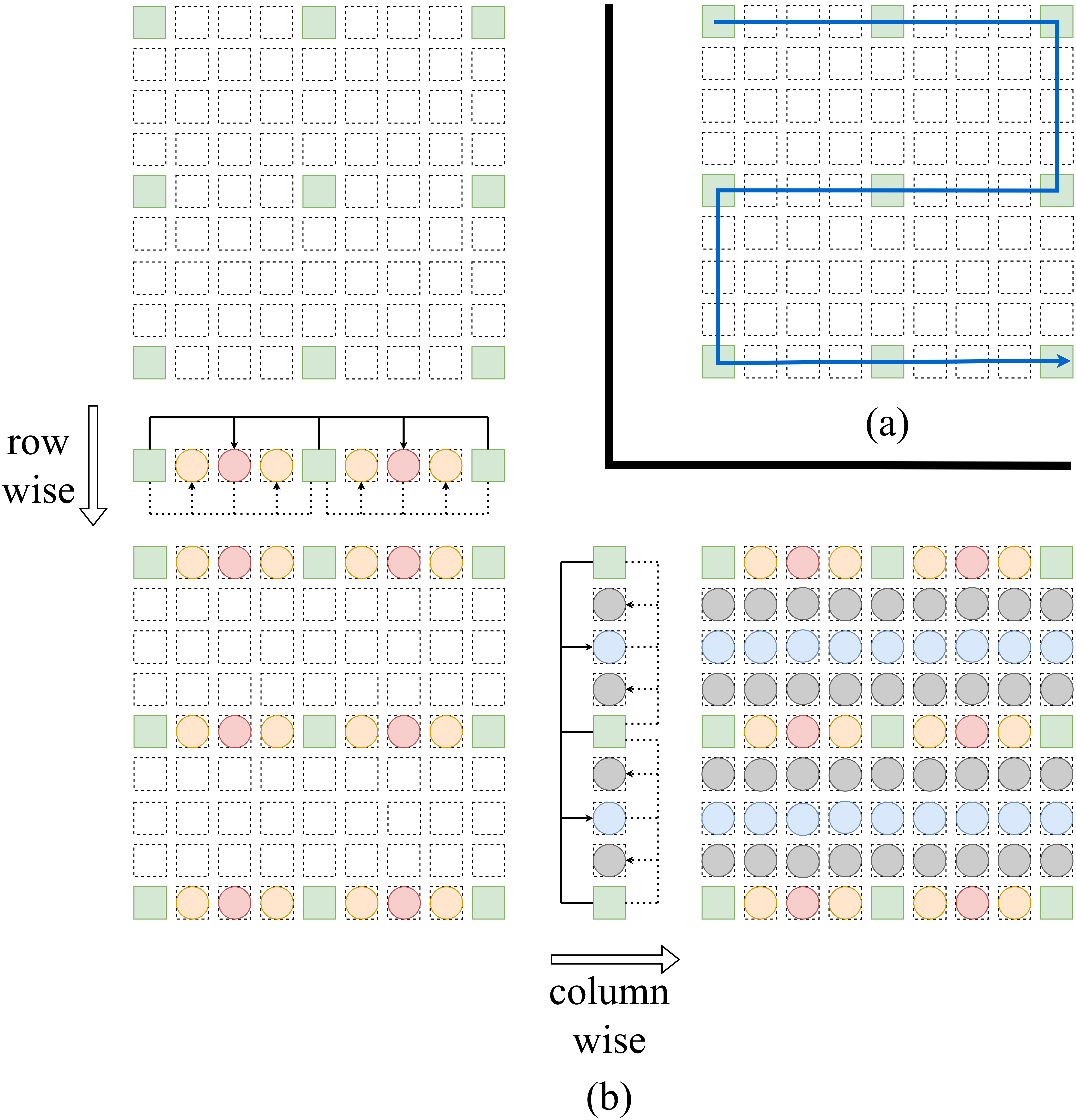}
\caption{(a) Snake order for compression and (b) multi-step reconstruction of a light field. Different colors indicate views reconstructed in different stages of a hierarchical process, i.e. (red, blue) circles mean first step and (yellow, grey) mean second step.
}
\label{fig:reconstruction}
\vspace{-0.4cm}
\end{figure}



\section{Subsampling and reconstruction for efficient light field compression}

In this section, we present different sub-sampling and reconstruction strategies for light field compression. 
We first select various sub-sampling strategies, then the sub-sampled views are encoded, and finally the light field is completed by view synthesis.

\subsection{Sub-sampling and Encoding}

With the classical two-plane representation of light fields, three basic view sub-sampling strategies are introduced, row, column and corners, as shown in Fig.~\ref{fig:subsampling}. Each one of them will have their corresponding reconstruction process, as discussed in Section~\ref{sec:Decoding}. We further investigate different sub-sampling densities (2x) and (4x). The remaining views after sub-sampling are scanned in snake order Fig.~\ref{fig:reconstruction}a and encoded with a video codec.

\begin{figure*}[!t]
\centering
\includegraphics[width=\linewidth]{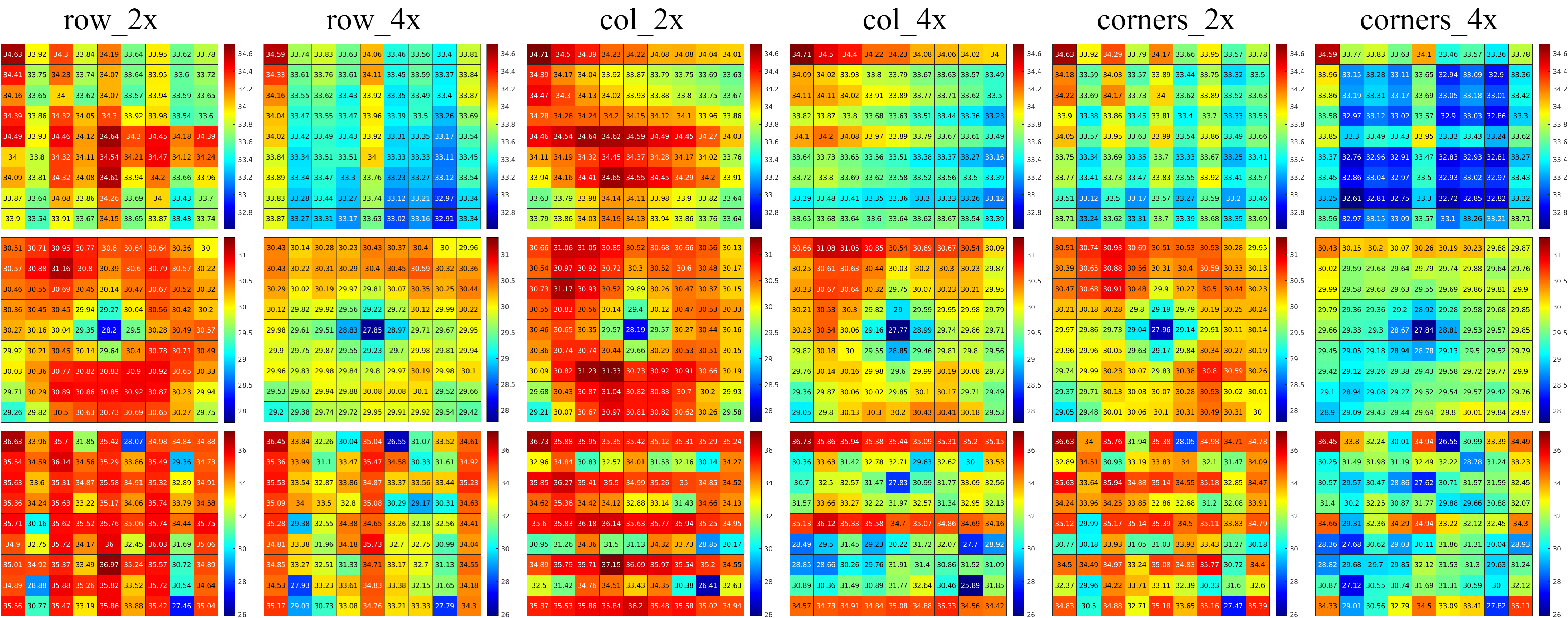}
\caption{PSNR scores after light field reconstruction with $QP=30$. Top, middle and bottom rows show results from HCI (\textbf{Bedroom}), Lytro (\textbf{Bee\_2}) and Stanford (\textbf{Lego Knights}) datasets using CycleLF~\cite{chen2020self}.}
\label{fig:matrix}
\vspace{-0.4cm}
\end{figure*}

\begin{figure}[!t]
\begin{tabular}{cc}
    \begin{minipage}{.48\linewidth}
        \centering
        \includegraphics[width=\linewidth]{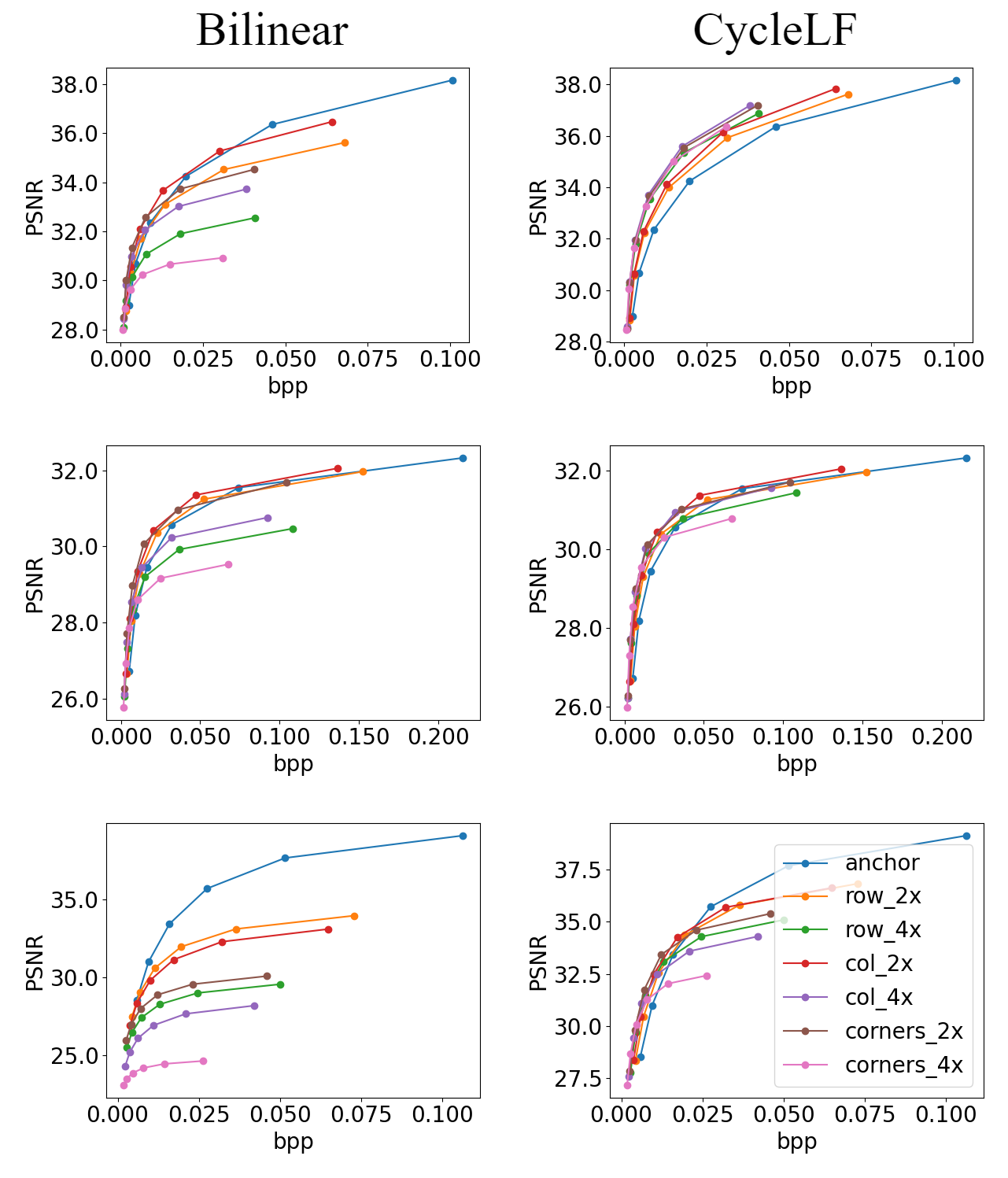}
        \caption{Rate-distortion results as PSNR over bit per pixel (bpp). Top row, middle row and bottom row are results on HCI (\textbf{Bedroom}), Lytro (\textbf{Bee\_2}) and Stanford (\textbf{Lego Knights}), respectively.}
        \label{fig:PSNR_BPP}
        \vspace{-0.5cm}
    \end{minipage}
    \hspace{0.2cm}
    \begin{minipage}{.48\linewidth}
        \centering
        \includegraphics[width=\linewidth]{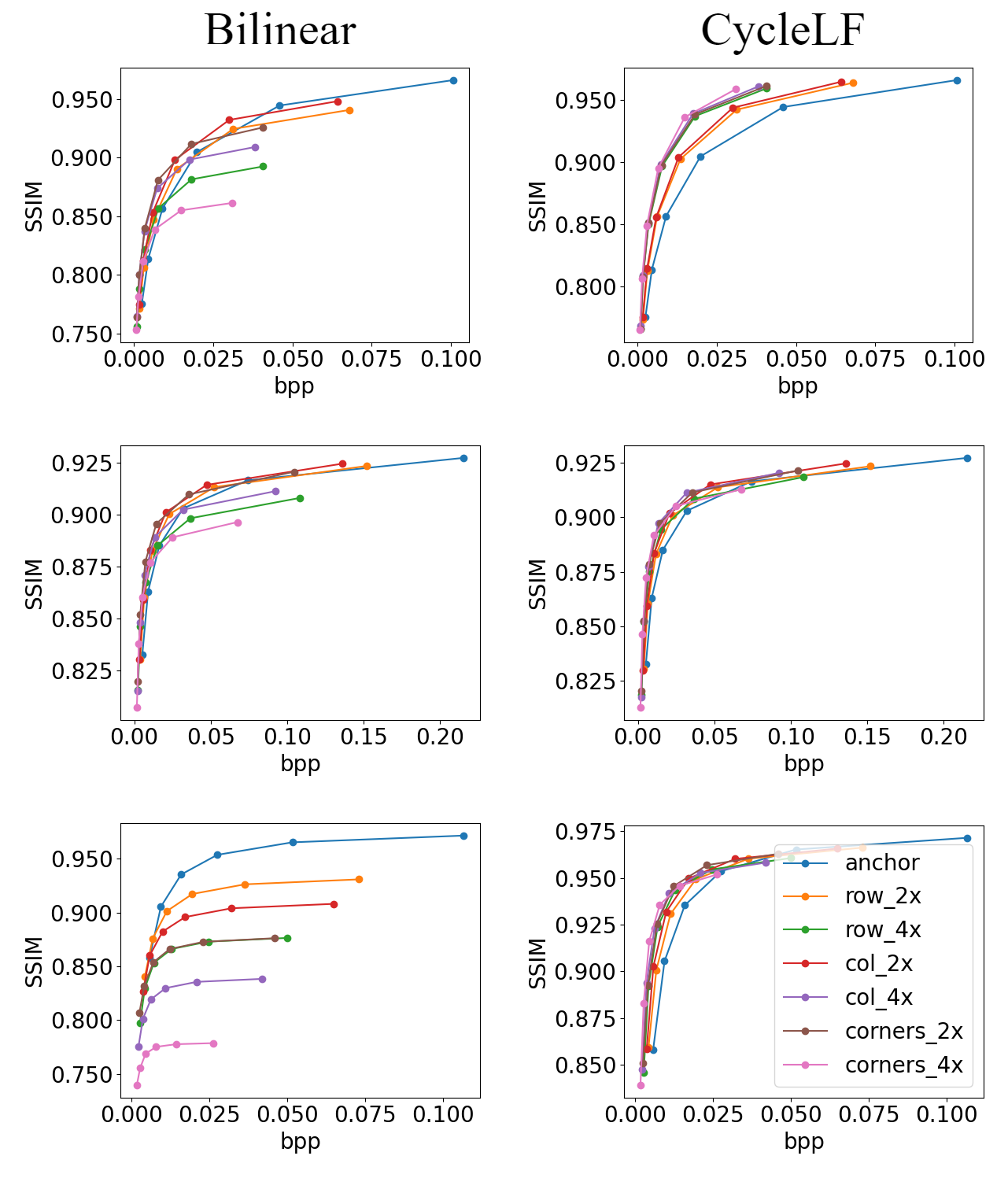}
        \caption{Rate-distortion results as Structural Similarity Index (SSIM) over bit per pixel (bpp). Top row, middle row and bottom row are results on HCI (\textbf{Bedroom}), Lytro (\textbf{Bee\_2}) and Stanford (\textbf{Lego Knights}), respectively.}
        \label{fig:SSIM_BPP}
        \vspace{-0.5cm}
    \end{minipage}
\end{tabular}

\end{figure}


\begin{figure*}
    \centering
    \begin{minipage}{.32\linewidth}
        \centering
        \includegraphics[width=\linewidth, scale=.15]{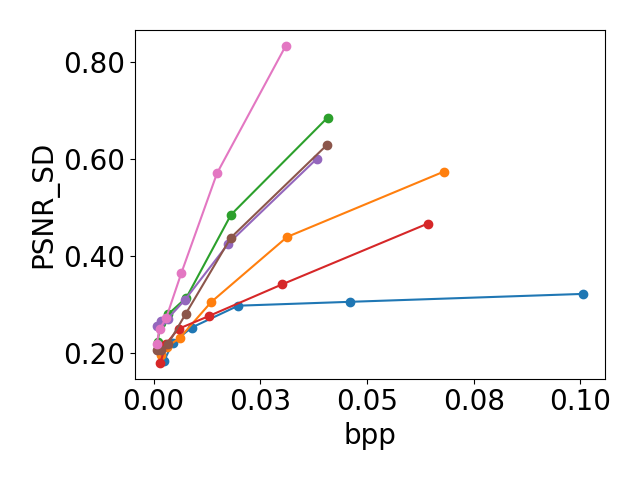}
    \end{minipage}
    \begin{minipage}{.32\linewidth}
        \centering
        \includegraphics[width=\linewidth, scale=.15]{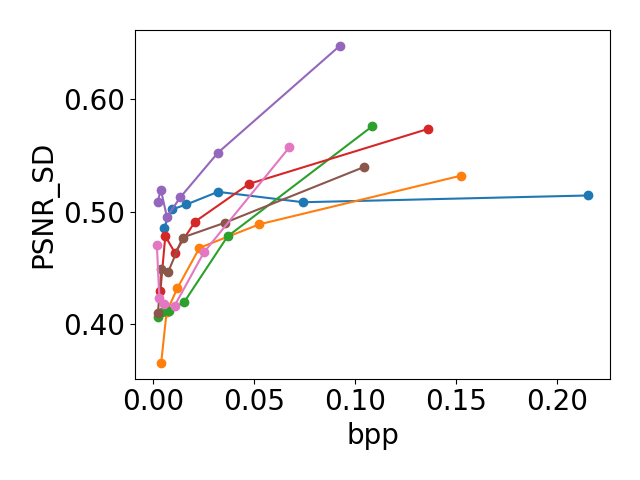}
    \end{minipage}
    \begin{minipage}{.32\linewidth}
        \centering
        \includegraphics[width=\linewidth, scale=.15]{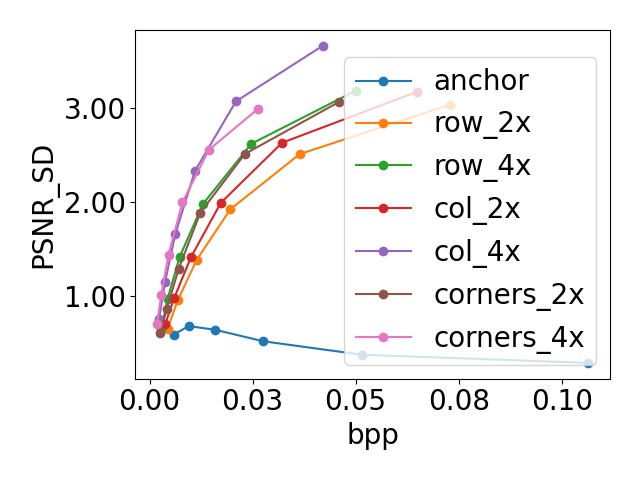}
    \end{minipage}
    \caption{Per-view standard deviations of PSNR results over bitrate. Left, middle and right for HCI (\textbf{Bedroom}), Lytro (\textbf{Bee\_2}) and Stanford (\textbf{Lego Knights}), respectively.}
	\label{fig:PSNR_SD_BPP}
	\vspace{-0.5cm}
\end{figure*}

\subsection{Decoding and Reconstruction}
\label{sec:Decoding}

After decoding at the receiver, we perform reconstruction by inverting the corresponding sub-sampling strategies. We take corners\_4x of a $9\times9$ light field as an example, as shown in Fig.~\ref{fig:reconstruction}b. The complete reconstruction process is a cascade of one row wise and one column wise view synthesis, while (4x) reconstruction is a cascade of two (2x) interpolations.
For row and column based sub-sampling, it is simple to apply row wise and column wise interpolation to complete the light field, accordingly. 

\setlength{\tabcolsep}{3pt}
\begin{table}[t]
\begin{tabular}{cc}
\begin{minipage}{.48\linewidth}
\begin{center}
\vspace{0.5cm}
\vspace{-0.3cm}
\begin{tabular}{cc|ccccc}
\noalign{\medskip}


& & \textbf{Bed} & \textbf{Bic} & \textbf{Her} & \textbf{Ori} & avg \\

\noalign{\smallskip} 
\hline 
\noalign{\smallskip} 
\multirow{2}{*}{row\_2x} & BD-PSNR & 0.68 & 0.60 & 0.51 & 0.47 & 0.57 \\
& BD-Rate & -24.8 & -22.6 & -21.4 & -22.3 & -22.8 \\


\noalign{\smallskip}
\hline
\noalign{\smallskip} 
\multirow{2}{*}{row\_4x} & BD-PSNR & 1.47 & 1.27 & 0.89 & 1.08 & 1.18 \\ 
& BD-Rate & -48.0 & -43.8 & -36.9 & -44.6 & -43.4 \\


\noalign{\smallskip}
\hline
\noalign{\smallskip}
\multirow{2}{*}{col\_2x} & BD-PSNR & 0.91 & 0.96 & 0.78 & 0.78 & 0.86 \\
& BD-Rate & -31.5 & -32.7 & -30.0 & -33.0 & -31.8 \\


\noalign{\smallskip}
\hline
\noalign{\smallskip} 
\multirow{2}{*}{col\_4x} & BD-PSNR & 1.79 & 1.72 & 1.26 & 1.54 & 1.58 \\
& BD-Rate & -53.8 & -51.6 & -46.1 & -55.0 & -51.6 \\


\noalign{\smallskip}
\hline
\noalign{\smallskip} 
\multirow{2}{*}{cors\_2x} & BD-PSNR & 1.66 & 1.56 & 1.24 & 1.26 & 1.43 \\
& BD-Rate & -51.6 & -48.4 & -44.8 & -49.3 & -48.5 \\

\noalign{\smallskip}
\hline
\noalign{\smallskip}

\multirow{2}{*}{cors\_4x} & BD-PSNR & 1.62 & 1.89 & 1.28 & 1.76 & 1.63 \\
& BD-Rate & -52.7 & -55.3 & -48.2 & -58.2 & -53.6 \\

\noalign{\smallskip}
\hline
\end{tabular}

\caption{BD-scores of CycleLF reconstruction on the HCI dataset, compared to baseline compression.}
\label{table:psnr_rd1}
\end{center}
\end{minipage}

\hspace{0.2cm}

\begin{minipage}{.48\linewidth}
\begin{center}
\vspace{0.5cm}
\vspace{-0.3cm}
\begin{tabular}{cc|ccccc}
\noalign{\medskip}


& & \textbf{Bee} & \textbf{Bik} & \textbf{Che} & \textbf{Des} & avg \\

\noalign{\smallskip} 
\hline 
\noalign{\smallskip} 
\multirow{2}{*}{row\_2x} & BD-PSNR & 0.28 & 0.47 & 0.58 & 0.30 & 0.49 \\
& BD-Rate & -18.1 & -20.8 & -28.9 & -19.7 & -24.8 \\


\noalign{\smallskip}
\hline
\noalign{\smallskip} 
\multirow{2}{*}{row\_4x} & BD-PSNR & 0.36 & 0.89 & 1.11 & 0.59 & 0.63 \\ 
& BD-Rate & -29.4 & -39.8 & -52.7 & -42.2 & -36.4 \\


\noalign{\smallskip}
\hline
\noalign{\smallskip}
\multirow{2}{*}{col\_2x} & BD-PSNR & 0.50 & 0.61 & 0.59 & 0.23 & 0.55 \\
& BD-Rate & -29.3 & -25.7 & -29.5 & -17.1 & -27.9 \\


\noalign{\smallskip}
\hline
\noalign{\smallskip} 
\multirow{2}{*}{col\_4x} & BD-PSNR & 0.68 & 0.89 & 1.10 & 0.34 & 0.48 \\
& BD-Rate & -41.6 & -41.3 & -52.0 & -36.6 & -36.6 \\


\noalign{\smallskip}
\hline
\noalign{\smallskip} 
\multirow{2}{*}{cors\_2x} & BD-PSNR & 0.63 & 1.05 & 1.18 & 0.52 & 0.97 \\
& BD-Rate & -39.1 & -43.7 & -54.1 & -40.7 & -47.3 \\

\noalign{\smallskip}
\hline
\noalign{\smallskip}

\multirow{2}{*}{cors\_4x} & BD-PSNR & 0.45 & 0.90 & 1.17 & 0.24 & 0.40 \\
& BD-Rate & -40.6 & -43.2 & -52.8 & -37.8 & -37.3 \\

\noalign{\smallskip}
\hline
\end{tabular}

\caption{BD-scores of CycleLF reconstruction on the Lytro dataset, compared to baseline compression.}
\label{table:psnr_rd2}
\end{center}
\end{minipage}

\end{tabular}
\end{table}

\section{Experiments}

In this section, we present the details of our experiments. All computations were performed on an Intel Core i7-6700k 4.0GHz CPU. 
To implement our pipeline, sub-sampled RGB views were converted into YUV 420 video using the open-source FFMPEG software~\cite{FFMPEG}. Then, these video files were encoded by an efficient video codec, HM (HEVC) 16.22~\cite{HEVC}. We used typical quantization parameters to vary bitrate and quality (QP: 20, 25, 30, 35, 40, 45).
As baseline comparison we also present the results of encoding the full light field without any sub-sampling and reconstruction, which is indicated as "anchor" in all the figures. 

We performed all experiments with synthetic light fields, Lytro images, and gantry-robot captured data to investigate different properties. The \textbf{Bedroom} light field is from the widely used synthetic HCI dataset. \textbf{Bee\_2} is extracted from the Lytro Illum dataset using a Lytro enhancement pipeline~\cite{Matysiak2018}. \textbf{LegoKnight} is from the Stanford dataset and pre-cropped to $512\times512$ to have similar spatial resolution to other light fields. This is a challenging light field due to its large disparity compared to others and extended textureless areas. From all light fields we used $9\times9$ views.


Figure~\ref{fig:matrix} shows the PSNR results after reconstruction as heatmaps, for 6 different strategies. These light fields were reconstructed by the state-of-the-art synthesis method CycleLF~\cite{chen2020self} with $QP=30$. We can recognize the sub-sampling patterns in these results, as directly decoded views have higher PSNR than interpolated views. These quality fluctuations are analyzed in more detail in Figure~\ref{fig:PSNR_SD_BPP}, where we show the standard deviations of PSNR results over the whole range of bitrates. We find that anchor encoding has the lowest fluctuation as expected. Fluctuations increase with the sub-sampling ratio as can also be expected, i.e. (4x) versions exhibiting highest fluctuations. While overall fluctuations are quite low for HCI and Lytro, they are significant for the relatively sparse Lego Knights. Thus, quality fluctuations have to be considered when applying view sub-sampling for light field compression.

Complete Rate-distortion (RD) results for PSNR and SSIM over bitrate are shown in Figure~\ref{fig:PSNR_BPP} and Figure~\ref{fig:SSIM_BPP}, respectively. 
All strategies with CycleLF outperform bilinear on all three datasets. All strategies with CycleLF reconstruction outperform anchor encoding on \textbf{Bedroom} and show equivalent performance on \textbf{Bee\_2}. Strategies with CycleLF reconstruction fall behind the anchor for \textbf{LegoKnight}, because of the large baseline of this light field affecting the performance of the reconstruction method. Meanwhile, regarding SSIM, CycleLF-based strategies consistently outperform the anchor. 


Bjontegaard metrics (BD-PSNR and BD-Rate)~\cite{wien2015high} are shown in Table~\ref{table:psnr_rd1} and Table~\ref{table:psnr_rd2} including a number of additional light fields. Compression with sub-sampling and CycleLF reconstruction consistently outperforms the anchor. Especially on the HCI dataset, the CycleLF-based method achieves an average BD-DSNR gain of 1.63dB and an average BD-DSNR bitrate saving of -53.6\% over anchor compression with the ``corner\_4x'' pattern.
Please note that we can't show BD-scores for the Stanford dataset as the large differences between the involved RD curves make this metric unsuitable for this case~\cite{wien2015high}.

\section{Conclusion}

In this paper we presented a comprehensive investigation about the influence of different view selection strategies on the light field compression task. To achieve this goal, we tested our complete pipeline including sub-sampling, encoding, decoding, and reconstruction with various strategies. Our results show that sub-sampling can improve compression efficiency, especially for dense light fields. Higher sub-samling can give more gain in these cases. However, fluctuations of output view quality have to be considered, which increase with the sub-sampling ratio.

\section*{Acknowledgments}

All authors are from the Trinity College Dublin, College Green, Ireland. Contact cheny5@tcd.ie for further questions about this work.
This publication has emanated from research conducted with the financial support of Science Foundation Ireland (SFI) under the Grant Number 15/RP/2776. We also gratefully acknowledge the support of NVIDIA Corporation with the donation of the Titan Xp GPU used for this research.
978-1-7281-9320-5/20/\$31.00 2020 European Union

\bibliographystyle{apalike}

\bibliography{imvip}

\begin{thebibliography}{}

\bibitem[FFM, ]{FFMPEG}
{FFMPEG}: A complete, cross-platform solution to record, convert and stream
  audio and video.
\newblock \url{https://www.ffmpeg.org/}.
\newblock accessed: 16-01-2022.

\bibitem[HEV, ]{HEVC}
{HM} reference software for high efficiency video coding {(HEVC)}.
\newblock \url{https://vcgit.hhi.fraunhofer.de/jvet/HM/}.
\newblock accessed: 16-01-2022.

\bibitem[Bakir et~al., 2020]{bakir2020light}
Bakir, N., Hamidouche, W., Fezza, S.~A., Samrouth, K., and D{\'e}forges, O.
  (2020).
\newblock Light field image coding using dual discriminator generative
  adversarial network and vvc temporal scalability.
\newblock In {\em 2020 IEEE International Conference on Multimedia and Expo
  (ICME)}, pages 1--6. IEEE.

\bibitem[Chen et~al., 2017]{chen2017light}
Chen, J., Hou, J., and Chau, L.-P. (2017).
\newblock Light field compression with disparity-guided sparse coding based on
  structural key views.
\newblock {\em IEEE Transactions on Image Processing}, 27(1):314--324.

\bibitem[Chen et~al., 2020a]{chen2020self}
Chen, Y., Alain, M., and Smolic, A. (2020a).
\newblock Self-supervised light field view synthesis using cycle consistency.
\newblock In {\em 2020 IEEE 22nd International Workshop on Multimedia Signal
  Processing (MMSP)}, pages 1--6. IEEE.

\bibitem[Chen et~al., 2020b]{chen2020study}
Chen, Y., Alain, M., and Smolic, A. (2020b).
\newblock A study of efficient light field subsampling and reconstruction
  strategies.
\newblock {\em arXiv preprint arXiv:2008.04694}.

\bibitem[Conti et~al., 2020]{conti2020dense}
Conti, C., Soares, L.~D., and Nunes, P. (2020).
\newblock Dense light field coding: A survey.
\newblock {\em IEEE Access}, 8:49244--49284.

\bibitem[Ebrahimi et~al., 2016]{ebrahimi2016jpeg}
Ebrahimi, T., Foessel, S., Pereira, F., and Schelkens, P. (2016).
\newblock Jpeg pleno: Toward an efficient representation of visual reality.
\newblock {\em Ieee Multimedia}, 23(4):14--20.

\bibitem[Jia et~al., 2018]{jia2018light}
Jia, C., Zhang, X., Wang, S., Wang, S., and Ma, S. (2018).
\newblock Light field image compression using generative adversarial
  network-based view synthesis.
\newblock {\em IEEE Journal on Emerging and Selected Topics in Circuits and
  Systems}, 9(1):177--189.

\bibitem[Jiang et~al., 2017a]{jiang2017light2}
Jiang, X., Le~Pendu, M., Farrugia, R.~A., and Guillemot, C. (2017a).
\newblock Light field compression with homography-based low-rank approximation.
\newblock {\em IEEE Journal of Selected Topics in Signal Processing},
  11(7):1132--1145.

\bibitem[Jiang et~al., 2017b]{jiang2017light}
Jiang, X., Le~Pendu, M., and Guillemot, C. (2017b).
\newblock Light field compression using depth image based view synthesis.
\newblock In {\em 2017 IEEE International Conference on Multimedia \& Expo
  Workshops (ICMEW)}, pages 19--24. IEEE.

\bibitem[Levoy and Hanrahan, 1996]{Levoy1996}
Levoy, M. and Hanrahan, P. (1996).
\newblock Light field rendering.
\newblock In {\em Proc. ACM SIGGRAPH}, pages 31--42.

\bibitem[Liu et~al., 2021]{liu2021view}
Liu, D., Huang, X., Zhan, W., Ai, L., Zheng, X., and Cheng, S. (2021).
\newblock View synthesis-based light field image compression using a generative
  adversarial network.
\newblock {\em Information Sciences}, 545:118--131.

\bibitem[Matysiak et~al., 2018]{Matysiak2018}
Matysiak, P., Grogan, M., Pendu, M.~L., Alain, M., and Smolic, A. (2018).
\newblock A pipeline for lenslet light field quality enhancement.
\newblock In {\em Proc. IEEE ICIP}, pages 639--643.

\bibitem[Singh and Rameshan, 2021]{singh2021learning}
Singh, M. and Rameshan, R.~M. (2021).
\newblock Learning-based practical light field image compression using a
  disparity-aware model.
\newblock {\em arXiv preprint arXiv:2106.11558}.

\bibitem[Viola et~al., 2018]{viola2018graph}
Viola, I., Maretic, H.~P., Frossard, P., and Ebrahimi, T. (2018).
\newblock A graph learning approach for light field image compression.
\newblock In {\em Applications of Digital Image Processing XLI}, volume 10752,
  page 107520E. International Society for Optics and Photonics.

\bibitem[Wafa et~al., 2021]{wafa2021learning}
Wafa, A., Pourazad, M.~T., and Nasiopoulos, P. (2021).
\newblock Learning-based light field view synthesis for efficient transmission
  and storage.
\newblock In {\em 2021 IEEE International Conference on Image Processing
  (ICIP)}, pages 354--358. IEEE.

\bibitem[Wien, 2015]{wien2015high}
Wien, M. (2015).
\newblock High efficiency video coding.
\newblock {\em Coding Tools and specification}, 24.

\bibitem[Zhao and Chen, 2017]{zhao2017light}
Zhao, S. and Chen, Z. (2017).
\newblock Light field image coding via linear approximation prior.
\newblock In {\em 2017 IEEE International Conference on Image Processing
  (ICIP)}, pages 4562--4566. IEEE.

\bibitem[Zhao et~al., 2018]{zhao2018light}
Zhao, Z., Wang, S., Jia, C., Zhang, X., Ma, S., and Yang, J. (2018).
\newblock Light field image compression based on deep learning.
\newblock In {\em 2018 IEEE International Conference on Multimedia and Expo
  (ICME)}, pages 1--6. IEEE.

\end{thebibliography}

\end{document}